# Symmetry-Driven Spin-Wave Gap Modulation in Nanolayered SrRuO$_3$/SrTiO$_3$ Heterostructures: Implications for Spintronic Applications

*Seung Gyo Jeong[†,#], Hyeonbeom Kim[‡,#], Sung Ju Hong[‡,#], Dongseok Suh[‡,*], Woo Seok Choi[†,*]*

[†]*Department of Physics, Sungkyunkwan University, Suwon 16419, Korea*

[‡]*Department of Energy Science, Sungkyunkwan University, Suwon 16419, Korea*

[*]Corresponding authors: energy.suh@skku.edu; choiws@skku.edu

KEYWORDS

SrRuO$_3$ heterostructure, anomalous Hall effect, Weyl semimetal, spin-wave gap, magnetic anisotropy control.

ABSTRACT

A strong correlation between magnetic interaction and topological symmetries leads to unconventional magneto-transport behavior. Weyl fermions induce topologically protected spin-momentum locking, which is closely related to spin-wave gap formation in magnetic crystals. Ferromagnetic SrRuO$_3$, regarded as a strong candidate for Weyl semimetal, inherently possesses a nonzero spin-wave gap owing to its strong magnetic anisotropy. In this paper, we propose a method to control the spin-wave dynamics by nanolayer designing of the SrRuO$_3$/SrTiO$_3$ superlattices. In particular, the six-unit-cell-thick SrRuO$_3$ layers within the superlattices undergo a phase transition in crystalline symmetry from orthorhombic to tetragonal, as the thickness of the SrTiO$_3$ layers is modulated with atomic-scale precision. Consequently, the magnetic anisotropy, anomalous Hall conductivity, and spin-wave gap could be systematically manipulated. Such customization of magnetic anisotropy via nanoscale heterostructuring offers a novel control knob to tailor the magnon excitation energy for future spintronic applications, including magnon waveguides and filters. Our nanolayer approach unveils the important correlation between the tunable lattice degrees of freedom and spin dynamics in topologically non-trivial magnetic materials.

INTRODUCTION

Magnetic interactions in topological states lead to the emergence of the exotic electromagnetic ground state of matter.[1-4] In particular, topological materials often exhibit characteristic magneto-transport behaviors, e.g., magnetoresistance (MR) and Hall effect, stemming from topological spin dynamics.[5, 6] For example, longitudinal negative MR originates from a chiral anomaly in Weyl semimetals.[7, 8] The anomalous Hall effect (AHE) can be understood as the integral of the Berry curvature over the occupied electronic states, and can be considered as a fingerprint of the existence of Weyl fermions.[9, 10] In general, Weyl semimetals have pairs of non-degenerate spin bands with band-touching near the Fermi energy in the momentum space. When a spin-exchange interaction is introduced in magnetically anisotropic Weyl systems, a gap opens between the two spin bands, leading to the formation of a spin-wave (magnon) gap.[11]

Ferromagnetic (FM) $SrRuO_3$ (SRO) is a candidate Weyl semimetal with a nonzero spin-wave gap.[9, 11-14] Chen *et al*. theoretically predicted the existence of Weyl nodes in SRO, which is manifested by the nonmonotonic AHE as a function of Fermi energy and magnetization ($M$) originating from the finite Berry curvature.[12] Recently, Itoh *et al*. observed the nonmonotonic temperature- ($T$-) dependent AHE and spin-wave gap formation in bulk SRO, which supported the existence of Weyl fermions along with magnon excitation.[11] Numerous transport and inelastic neutron scattering experiments further examined the effect of the Weyl fermions and magnon dynamics in SRO.[11, 13-16]

The formation of the spin-wave gap and its dynamics in SRO are closely related to the magnetic anisotropy of SRO.[17, 18, 19] In principle, a strong magnetic anisotropy is necessary to develop the nonzero spin-wave gap in FM systems. For example, general FM materials with negligible magnetic anisotropy, such as $La_{0.8}Sr_{0.2}MnO_3$, exhibit gapless spin-wave dispersions.[17] The magnon gaps, when stabilized, are useful for realizing spintronic devices with fast, energy-efficient performances, for example, in magnon waveguides and filters.[20] Most previous studies on magnon gaps have focused on micro-structured waveguides (magnonic crystals) for the excitation and modulation of spin-wave gaps.[20-23] In real crystals, however, the formation of the nonzero spin-wave gap is highly unusual, posing a challenge for its viable application. In the case of SRO, strong anisotropy, combined with Weyl states, topologically locks the spin momentum. This stabilizes the spin-wave gap and provides an opportunity to customize the magnetic anisotropy and magnon gap.

Deliberate nanoscale precision heterostructuring is a viable solution for modulating the magnetic anisotropy and spin-wave gap of SRO. Through heterostructuring, the lattice degree of freedom, including the crystalline symmetry, has become an accessible control parameter to customize the physical properties of SRO heterostructures.[18, 24-27] In bulk, SRO has a $GdFeO_3$-type orthorhombic crystal symmetry with distorted $RuO_6$ octahedra, which is closely associated with its magnetic ground state. Various SRO heterostructures have been proposed to adjust the tilt of $RuO_6$ octahedra and its influence on FM behavior. Particularly, a $SrRuO_3/SrTiO_3$ superlattice (SRO/STO SL) is an ideal prototype for selectively controlling the crystalline symmetry and magnetic anisotropy of SRO.[28-30]

In this study, we demonstrate the customization of the spin-wave gap in SRO through artificial crystalline symmetry control. We fabricated SRO/STO SLs using pulsed laser epitaxy with atomic-scale precision. We fixed the thickness of the SRO layers as six-unit cells (u.c.), whose crystalline symmetry could be systematically modified from an orthorhombic ($SRO_o$) to a tetragonal ($SRO_t$) (the subscripts "$o$" and "$t$" represent "orthorhombic" and "tetragonal", respectively.) upon increasing the STO u.c. layer thickness from two to eight u.c. within the SL. The artificial crystalline symmetry modification provides controllability of the magnetic anisotropy energy ($E_k$). Figure 1(a) shows the strategy for customizing the spin-wave gap of the SRO heterostructure. It has been reported that the magneto-crystalline anisotropy constant ($K$) of $SRO_t$ ($K_t$) was larger than that of $SRO_o$ ($K_o$)[31, 32] with a simultaneous rotation in the magnetic easy axis ($\theta_{ea}$, arrows in Figure 1(a)) from a finite, oblique angle in $SRO_o$ to an out-of-plane direction in $SRO_t$. The collectively modulated $E_k$ (~$K\cos^2\theta_{ea}$) influences the nonmonotonic $T$-dependent AHE, indicating the evolution of the spin-wave gap ($E_g$) and its dynamics.

Atomically designed SRO/STO SLs were synthesized using pulsed laser epitaxy to realize a phase transition in the crystalline symmetry of SRO, as schematically shown in Figure 2(a). Figures 2(b)-2(d) show the precise manipulation of the atomic u.c. layers of the SRO/STO SL series on a single-crystalline (001) STO substrate. [(SRO)$_6$|(STO)$_y$]$_{10}$ indicates that six-u.c. layers of SRO and $y$-u.c. layers of STO are repeated ten times along the growth direction. X-ray diffraction (XRD) $\theta$-$2\theta$ measurements (Figure 2(b)) and reciprocal space maps (Figure 2(c)) show clear SL peaks that correspond to the nanoscale periodicity of each SL under the fully strained state. Figure 2(d) shows the high-angle annular dark field-scanning transmission electron microscopy (HAADF-STEM) images indicating the well-ordered atomic columns of SLs (The images were reproduced from Ref. 29). Figure 1(b) illustrates the RuO$_6$ octahedral tilt angle ($\theta_{tilt}$) in the scheme of the pseudocubic SRO unit cell, defining the crystalline symmetry. In general, $SRO_o$ shows a finite $\theta_{tilt}$, whereas $SRO_t$ suppresses $\theta_{tilt}$ to 0º. The orthorhombicity, defined as $a_o/b_o$ (see the inset of Figure 1(b) for the orthorhombic unit cell), was estimated from off-axis XRD $\theta$-$2\theta$ scans of the SLs around the (204) STO plane (Figure S1).[33] Figure 1(b) shows the $y$-dependent $a_o/b_o$ for the [(SRO)$_6$|(STO)$_y$]$_{10}$ SLs with the $y$-dependent structural phase

transition from the orthorhombic to the tetragonal symmetry. Based on $a_o/b_o$, $\theta_{tilt}$ was estimated as $\theta_{tilt} = \cos^{-1}(b_o/a_o)$, assuming that the Ru-O-Ru bond length change is negligible.[34] Furthermore, $\theta_{tilt} = \sim 6°$, which is relatively suppressed compared to that of the bulk (~10°), was obtained for the SL for which $y = 2$. As $y$ increases, $\theta_{tilt}$ decreases and becomes zero for $y \geq 6$ (*i.e.*, $a_o = b_o$). It can be inferred that, when thick enough, the adjacent cubic layers of STO within the SL restrain $\theta_{tilt}$ of the SRO layer, thereby stabilizing the tetragonal symmetry for the SLs.[29] Meanwhile, the *T*-dependent resistivities ($\rho_{xx}$) of the SLs exhibit a typical kink near the common FM phase transition temperature ($T_c$) of ~130 K, as shown in Figure S2 (also see Supplemental Note 1). This indicates that the itinerant ferromagnetic nature of SRO is well-preserved even for the atomically thin SRO layers in our SLs, irrespective of $y$.

The application of the magnetic field (*H*) allows us to estimate the crystalline-symmetry-dependent change in transport behavior (Figure 3). This magneto-transport behavior is rather unexpected because the transport behavior was relatively consistent in the absence of the external *H* (Figure S2). First, *H*-dependent MR (MR (*H*) = [$\rho_{xx}$ (*H*) – $\rho_{xx}$ (0)]/$\rho_{xx}$ (0)) shows a strong structural dependence. The hysteresis behavior of MR (*H*) curves shows the typical FM feature of SRO, with the peaks corresponding to the coercive field ($H_c$) (Figure 3(a)). The $H_c$ is systematically enhanced with increasing *y*, whereas the negative MR value becomes less negative. The larger negative MR values of SRO$_o$ than those of SRO$_t$ might partially result from the enhanced saturation magnetization ($M_s$) in SRO$_o$ (Figure S3), which will be discussed later in more detail. Second, *H*-dependent Hall resistivity ($\rho_{xy}$ (*H*)) further manifests a crystalline symmetry dependence. (Figure 3(b)). In general, $\rho_{xy}$ of SRO is defined by

$$\rho_{xy} = \rho_{\text{OHE}} + \rho_{\text{AHE}} = R_0 H + R_s \mu_0 M, \tag{1}$$

where $\rho_{OHE}$, $\rho_{AHE}$, $R_0$, $R_s$, and $\mu_0$ are the ordinary Hall effect (OHE) resistivity, AHE resistivity, OHE coefficient, AHE coefficient, and vacuum permeability, respectively. $\rho_{AHE}$ is closely related to spin-ordered electronic transports for SRO in the FM phase. The $H_c$, defined based on the magnetic hysteresis of $\rho_{xy}(H)$, increases systematically with increasing $y$, which is consistently the case for both MR$(H)$ and $M(H)$ (Figure 3(c)).

More distinct crystalline symmetry dependences of the SRO SLs are captured in terms of $M_s$ and $\rho_{AHE}$ (Figure 3(d)). We extracted $\rho_{AHE}$ (4 T) by accurately excluding the contribution of OHE from $\theta_H$-dependent $\rho_{xy}$ (Figure S4(a)).[35] The following equation was used to obtain $\rho_{OHE}$:

$$\Delta\rho_{xy}(\theta_H) = \rho_{xy}(\theta_H) - \rho_{xy}(0\ \text{T}) = R_0 H \cos\theta_H + ((d\rho_{AHE})/(dM))\chi H \cos(\theta_{ea} - \theta_H), \qquad (2)$$

where $\chi$ is the magnetic susceptibility and $\theta_H$ is the angle of $H$ from the film normal direction. The conventional linear-fit-based separation of the OHE contribution (Figure S4(b)) shows consistent crystalline symmetry dependence. The resultant $\rho_{AHE}$ (4 T) shows an abrupt jump across the structural phase transition, wherein the SLs with SRO$_t$ exhibit significantly less negative values. In contrast, we note that the $M_s$ values are larger for the SLs with SRO$_o$ than for those with SRO$_t$ (Figure S3), indicating that $\rho_{AHE}$ cannot be obtained based on the simple linear proportionality of $M$. This implies that physical origins other than magnetism are playing a role in determining $\rho_{xy}$ of the SLs.

Figure 4(a) shows nonmonotonic $\rho_{xy}(T)$ curves corresponding to the SRO/STO SLs, possibly originating from the intrinsic Berry curvature in SRO layers.[11, 13] Above $T_c$, the SLs exhibit identical positive $\rho_{xy}(T)$ values. With decreasing $T$, $\rho_{xy}(T)$ decreases with an inflection point near $T_c$ and a sign change below $T_c$. The $T$-dependent behavior again manifests that $\rho_{xy}$ cannot be understood from $M$ alone, as the $T$-dependent magnetization ($M(T)$) curves of the SLs exhibit a monotonic increase below $T_c$ (Figure S5). Haham *et al.* proposed that the normalized $R_s$ for SRO single films with various thicknesses can be scaled with respect to normalized $\rho_{xx}$ at a finite $H$ (Figure S6(c)). Here, the AHE was explained base on both the contributions of the extrinsic side-jump scattering and the intrinsic Berry phase effect.[15] We similarly extracted normalized $R_s$ of the SLs and plotted it as a function of the normalized $\rho_{xx}$ ($T$) at $H$ = 4 T (Figure S6). The normalized $R_s$ curve of the SRO/STO SL consistently scales with that obtained from the SRO single film, supporting the arguments by Haham *et al.* Thus, the result indicates that an intrinsic Berry phase effect contributes to the AHE of the SLs.

The AHE conductivity ($\sigma_{AHE}$) further depicts the crystalline-symmetry-dependent AHE of the SRO/STO SLs. Figure 4(b) shows the $T$-dependent $\sigma_{AHE}$ with different $y$ values, estimated using $\sigma_{AHE}$ ($T$) ~ $-\rho_{AHE}$ ($T$)/$\rho^2_{xx}$ ($T$) with the $H$-field of 4 T (Figure S6). Notably, at low $T$, $\sigma_{AHE}$ of SRO$_o$ is significantly larger than that of SRO$_t$. Such distinct differences in $\sigma_{AHE}$ with respect to crystalline symmetry change have not yet been observed.[9, 36, 37] We further note that $\sigma_{AHE}$ of SRO leads to the estimation of the spin-wave gap ($E_g$).[11, 14, 37] Furthermore, $E_g$ can be estimated as

$$E_g\ (T) = \frac{a(M\ (T)/M\ (5\ \text{K}))}{1 + b(M\ (T)/M\ (5\ \text{K}))(\sigma_{AHE}\ (T)/\sigma_0)}\ , \tag{3}$$

where $a$ = 3.2 meV and $b$ = –9.5, which are adopted from the inelastic neutron scattering result.[11] In addition, $\sigma_0 = e^2/ha_{pc} = 9.9 \times 10^2$ S cm$^{-1}$ is a normalization factor for $\sigma_{AHE}$, where $e$, $h$, and $a_{pc}$ are the elementary charge, Planck constant, and pseudocubic lattice constant of SRO, respectively. Figure 4(c) shows the nonmonotonic $T$-dependent $E_g$ of the SLs for different $y$ values. While the $E_g$ values obtained in this study are in the same order to that in previous studies, it reveals a clear crystalline-symmetry dependence. The inset of Figure 4(c) shows the $y$-dependent $E_g$ at 5 K. SRO$_t$ in the SLs corresponding to $y$ = 6 and 8 exhibits a larger $E_g$ value compared to the SRO$_o$ in the SLs with $y = 2$ and 4. Notably, a 42% enhancement of $E_g$ within the SLs is achieved. We note that $E_g$ of the SRO was estimated as 1 – 2 meV from neutron scattering experiments and 1 meV from ferromagnetic resonance experiments, depending on the sample preparation (either single crystal or powder), crystal orientation, and/or dimensionality. [11, 14, 38] $E_g$ of the FM phase can be generally described using $E_g \sim E_k<S^z>$, where $<S^z>$ is the average spin moment. Thus, a change in $M_s$ can affect $E_g$ through $<S^z>$; however, the crystalline-symmetry-dependent variation of $M_s$ is less than ~10% (Figure S3). Consequently, the crystalline-symmetry-dependent $E_k$ is thought to dominantly influence the spin-wave gap values of the SRO/STO SLs. Previous studies show a substantial difference in the anisotropy field (~$2K/M_s$) between the tetragonal (~12.0 T) and orthorhombic SRO (~7.2 T), justifying the modification in $E_k$.[31, 32, 38] This in turn can explain the significant enhancement of $E_g$ as the SRO/STOs become tetragonal with increasing $y$.

In the meantime, we noticed that the $E_g$ value of $y = 4$ SL is larger than that of the $y = 2$ SL, even though both SLs have the same crystalline symmetry of SRO$_o$. We measured $\theta_H$-dependent $\rho_{xy}$ of the SLs to understand the difference, as shown in Figure 4(d). We used $H = 4$ T, which is smaller than the strong magnetic anisotropy field of SRO that is of the order of 10 T.[30] The clockwise and anticlockwise $H$ rotation measurements of $\rho_{xy}$ ($\theta_H$) exhibit an abrupt increase and decrease every 180º, indicating the $M$ reversal of FM SRO. Furthermore, hysteresis is observed depending on the direction of the field rotations, indicating the FM nature of the SRO layer. $\theta_{ea}$ of the SRO/STO SLs can be identified as the central angle of hysteresis.[18] The bulk SRO crystal has $\theta_{ea} = 30º$, which was proposed to be closely related to its $\theta_{tilt} = $ ~10º.[18, 39, 40] Among our SRO/STO SLs, the $y = 2$ SL exhibits $\theta_{tilt} = $ ~6º and $\theta_{ea} = 17º$, which is suppressed compared to that of the bulk, as shown in Figure 4(d). When $y = 4$, $\theta_{tilt} = $ ~3º and $\theta_{ea} = 0º$, in accordance with the positive relation between $\theta_{tilt}$ and $\theta_{ea}$. The $y = 6$ and 8 SLs possess SRO$_t$ with $\theta_{tilt} = 0°$ and $\theta_{ea} = 0°$. The difference in $\theta_{ea}$ between the $y = 2$ and 4 SLs can account for the subtle difference in $E_k$ through $\cos^2\theta_{ea}$, and for the resultant ~10% difference in $E_g$, even for the same crystalline symmetry. These results consistently reflect the strong correlation between $\sigma_{AHE}$, $E_g$, and $E_k$ of the SRO/STO SLs.

CONCLUSION

In summary, the artificial crystalline symmetry customization was achieved for the SRO layer within nanolayered SRO/STO heterostructure, to systematically modulate the magnetic anisotropy and resultant evolution of spin-wave gap. The magneto-transport measurements demonstrated that the magnon gap could be substantially enhanced by changing the crystalline symmetry of the SRO from an orthorhombic to a tetragonal structure. Our approach demonstrates the atomic-scale engineering of the Berry curvature and spin-wave dynamics in artificial magnetic crystal for next-generation spintronic applications.

METHODS

**Synthesis of artificial superlattice and characterization of the periodic lattice structure.**
Nanolayered [(SRO)$_6$|(STO)$_y$]$_{10}$ SLs with six-u.c. layers of SRO and $y$-u.c. layers of STO were synthesized using pulsed laser epitaxy on (001) STO substrates, as schematically shown in Figure 2(a). All SLs are repeated ten times along the growth direction, with the last layer ending

with the STO layer. Both SRO and STO layers were deposited at 750 ºC and 100 mTorr of oxygen partial pressure, to simultaneously ensure the stoichiometric conditions for both of the materials. We ablated stoichiometric ceramic targets using a KrF laser (248 nm, IPEX868, Lightmachinery) with a laser fluence of 1.5 J cm$^{-2}$ and a repetition rate of 5 Hz. We systematically controlled the number of atomic u.c. in the SLs by employing a customized automatic laser pulse control system programmed using LabVIEW. Structural characterization of the SLs was carried out using high-resolution X-ray diffraction (HRXRD) via Rigaku Smartlab and a PANalytical X'Pert X-Ray Diffractometer. We estimated the thickness of the SL period using Bragg's law as

$$\Lambda = \frac{\lambda}{2}(sin\:\theta_n - \:sin\:\theta_{n-1})^{-1},$$

where $\Lambda$, $n$, $\lambda$, and $\theta_n$ are the period thickness, SL peak order, X-ray wavelength, and $n$th-order SL peak position, respectively. All the SLs exhibit a small thickness deviation, below 1 u.c. (~0.4 nm), corresponding to the atomic step-size of the substrate.

**Magnetization and electrical transport characterization.** $M\:(T)$ and $M\:(H)$ were measured using a Magnetic Property Measurement System (MPMS, Quantum Design). The $M\:(T)$ measurements were performed from 300 to 2 K under a magnetic field of 100 Oe along the out-of-plane direction of the SLs. The $M\:(H)$ curves were obtained at 5 K with the magnetic field along the out-of-plane direction. Electrical transport measurements were carried out using the van der Pauw geometry with an excitation current of 100 $\mu$A. Note that measured Hall voltages with two orthogonal current directions show consistent FM hysteresis, indicating the absence of in-plane anisotropy.[41] Various $T$ (300 to 2 K), $H$ (0 to 4 T), and tilt angles (0 to 360º) were mapped for both $\rho_{xx}$ and $\rho_{xy}$ using a Physical Property Measurement System (PPMS, Quantum Design).

FIGURES

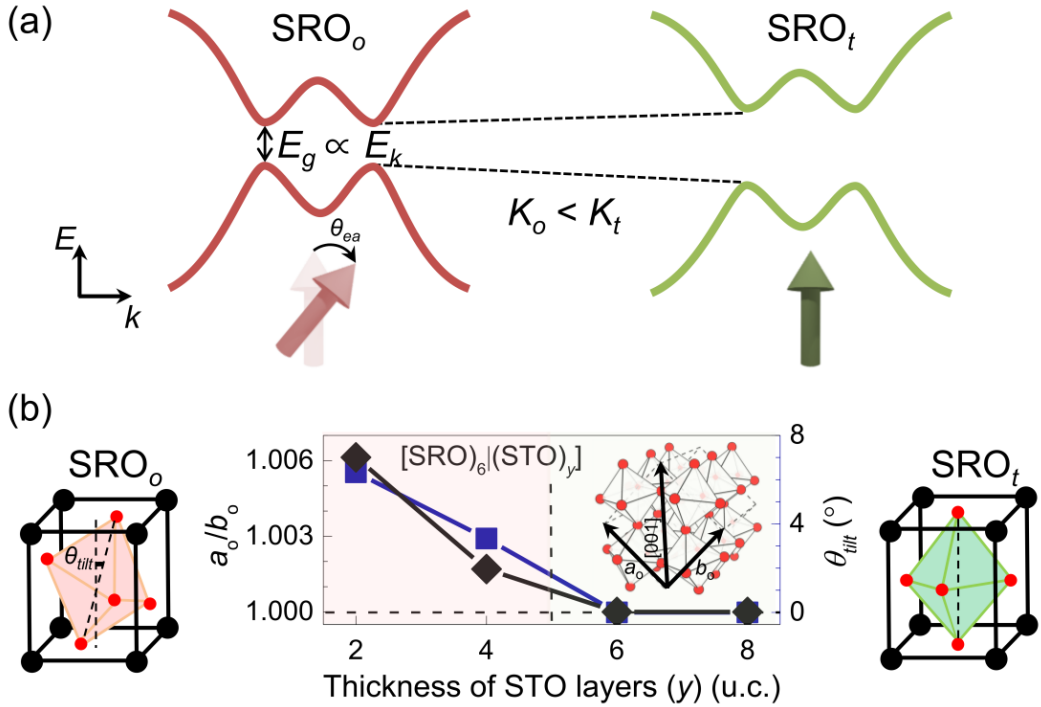

**Figure 1.** Customization of spin-wave gap of SRO in nanolayered SRO/STO SLs. (a) Key features of spin-wave gap ($E_g$) control via deliberately modified magnetic anisotropy. Crystalline-symmetry-dependent $K$ and $\theta_{ea}$ of SRO collectively determine the $E_g$ value. The direction of the arrow indicates $\theta_{ea}$ of SRO. (b) $a_o/b_o$ (black symbols) and $\theta_{tilt}$ (blue symbols) of the six-u.c. SRO SLs were systematically modulated based on the atomically controlled thickness of STO layers ($y$), accompanied with the structural phase transition. The red (green) region indicates the orthorhombic (tetragonal) structure of the SLs. The schematics of the pseudocubic u.c. of SRO represent the structural phase transition from the $SRO_o$ (left panel) to the $SRO_t$ (right panel). The inset schematically shows orthorhombic distortions ($a_o/b_o$) extracted from the lattice parameters of the orthorhombic u.c..

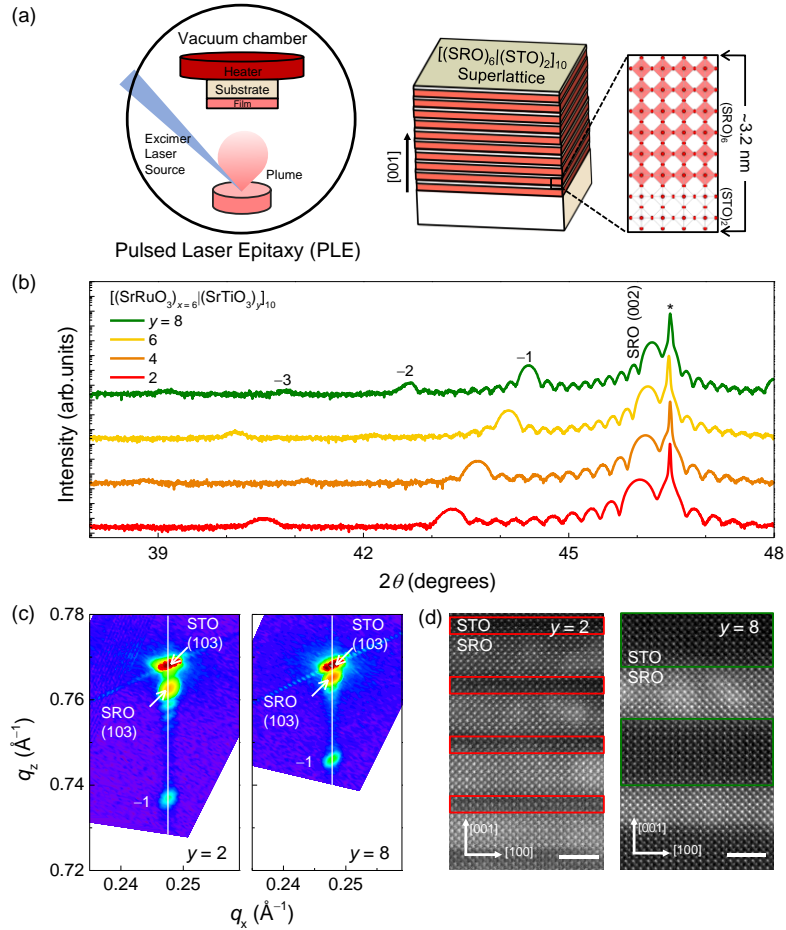

**Figure 2.** Nanolayered SRO/STO SLs with different $y$ values. (a) Schematic representation of the pulsed laser epitaxy system (left panel) to synthesize nanolayered SRO/STO SLs, for example, $y = 2$ SL (right panel). (b) XRD $\theta$-$2\theta$ scans, around the STO (002) Bragg reflections, are shown for the nanolayered $[(SRO)_6|(STO)_y]_{10}$ SLs with different $y = 2, 4, 6,$ and $8$. The Bragg peaks of the SL ($-n$) indicate the well-defined and controlled periodicity. The asterisk (*) indicates the STO (002) substrate peaks. (c) Reciprocal space maps of the $y = 2$ and 8 SLs were obtained around the STO (103) Bragg reflection. The vertical lines manifest the coherently strained SL layers. (d) Cross-sectional HAADF-STEM results are shown for the SLs with $y = 2$ (left panel) and 8 (right panel), redrawn from a previous study.[29] The bright layers indicate the SRO layers. The red and green rectangle indicate the modulated STO layers for the $y = 2$ and 8 SLs. The scale bars denote 2 nm.

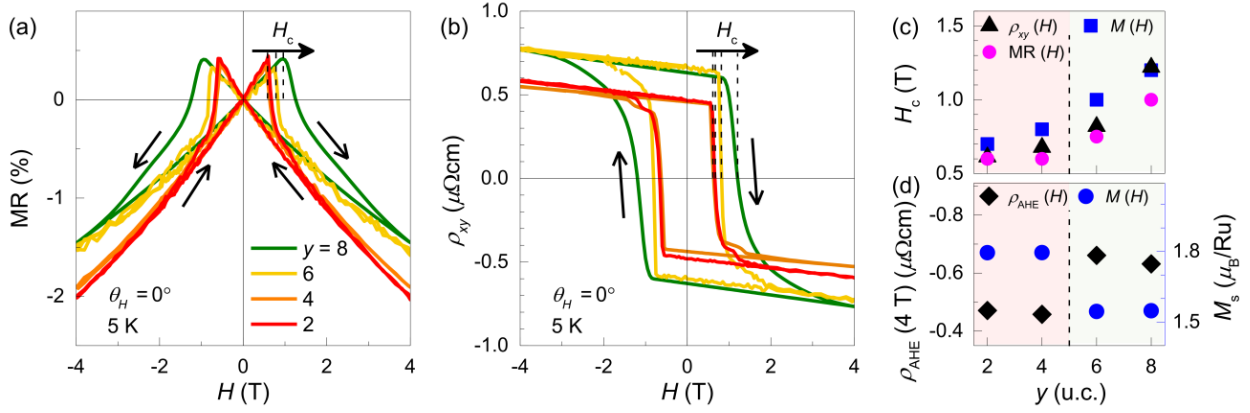

**Figure 3.** Structural dependence of magneto-transports in SRO/STO SLs. (a) $H$-dependent MR ($[\rho_{xx}(H) - \rho_{xx}(0\,\text{T})]/\rho_{xx}(0\,\text{T})$) and (b) $\rho_{xy}(H)$ curves of SLs were measured at 5 K and with the out-of-plane magnetic field ($\theta_H = 0°$). The arrows indicate the $H$-field sweeping direction. (c) The $\rho_{xy}(H)$, MR($H$), and $M(H)$ curves of SLs consistently exhibit an increase in $H_c$ with increasing $y$. (d) $\rho_{\text{AHE}}$ and $M_s$ were obtained at 5 K with $H$ of 4 T. The red (green) region indicates the SRO$_o$ (SRO$_t$).

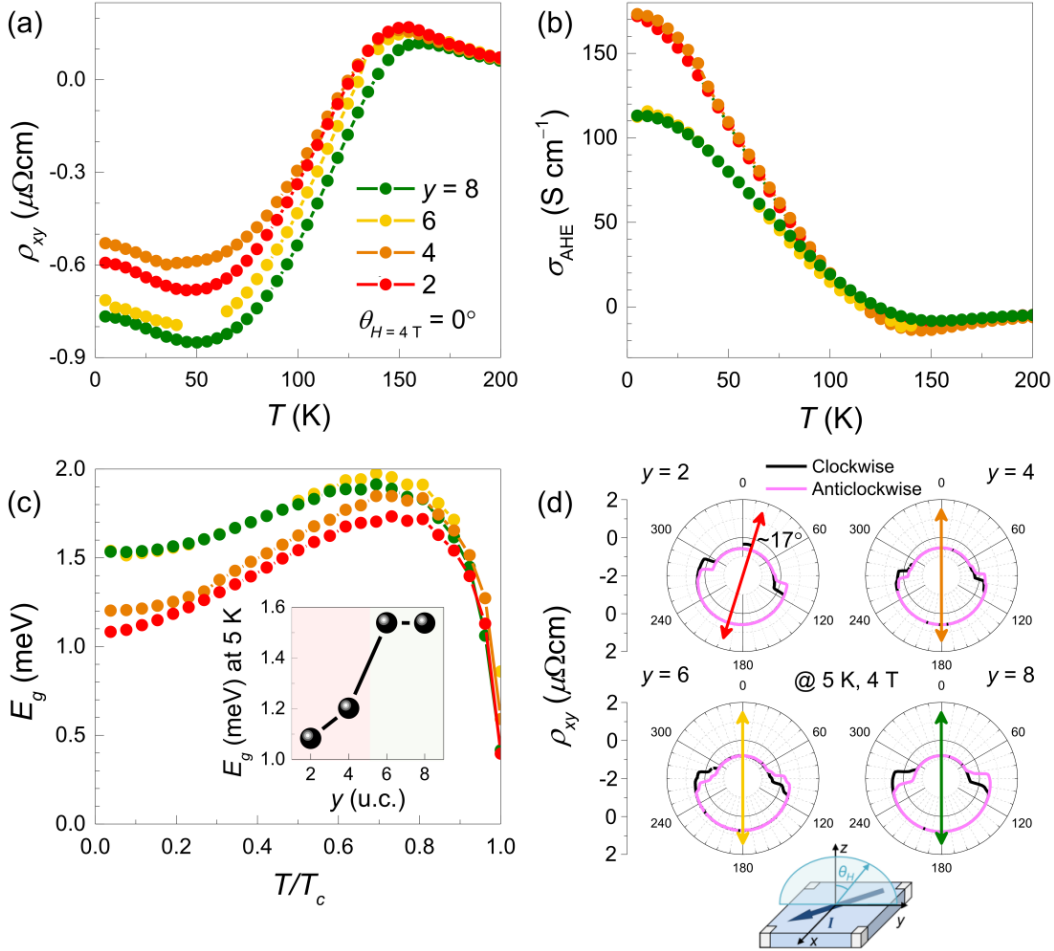

**Figure 4.** Symmetry-driven AHE with a modulated spin-wave gap of SRO/STO SLs. (a) nonmonotonic $\rho_{xy}$ (*T*) of the SRO SLs with different *y* shows a crystalline symmetry dependence. $\rho_{xy}$ (*T*) was measured at 5 K with an out-of-plane *H* of 4 T (b) $\sigma_{AHE}$ (*T*) of $SRO_o$ and $SRO_t$ show a strong structural dependence. (c) $E_g$ (*T*) can be modulated via the artificial crystalline symmetry of SRO. Inset in (c) indicates $E_g$ at 5 K for different *y* values. (d) The $\rho_{xy}$ ($\theta_H$) curves at 5 K with the *H* = 4 T show the *y*-dependent $\theta_{ea}$ of the SRO SLs. The measurement was performed with (anti-) clockwise field rotation. The *y* = 2 SL shows a distinct $\theta_{ea}$ of ~17°, whereas $\theta_{ea}$s for *y* = 4, 6, and 8 are zero. The bottom schematic shows the configuration employed for $\rho_{xy}$ ($\theta_H$) measurements.

## ASSOCIATED CONTENT

The Supporting Information is available free of charge at

Details of electrical, magnetic, and structural characterizations, crystalline symmetry-dependent carrier concentration, resultant $M_s$, separation of OHE and AHE using $\theta_H$-dependent analysis, and scaling behavior with $R_s$ ($\rho_{xx}$), nanolayered SRO/STO SLs with different $y$ (PDF).

## AUTHOR INFORMATION

### Corresponding Authors

**Woo Seok Choi** − *Department of Physics, Sungkyunkwan University, Suwon 16419, Korea*; https://orcid.org/0000-0002-2872-6191; E-mail: choiws@skku.edu

**Dongseok Suh** − *Department of Energy Science, Sungkyunkwan University, Suwon 16419, Korea*; https://orcid.org/0000-0002-0392-3391; E-mail: energy.suh@skku.edu

### Author Contributions

[#]These authors contributed equally to this work. S.G.J. and W.S.C. designed and analyzed the experiments. S.G.J. synthesized the superlattice and characterized the structural symmetry. H.K., S.J.H., and D.S. performed the electrical transport measurements. S.G.J., H.K., and S.J.H. wrote the initial manuscripts. W.S.C. and D.S. led the project. All authors prepared the manuscript together.

### Notes

The authors declare no competing financial interest.

ACKNOWLEDGMENT

This work was supported by the Basic Science Research Programs through the National Research Foundation of Korea (NRF) (NRF-2019R1A2B5B02004546, NRF-2019R1I1A1A01058123, and NRF-2019R1A2B5B02070536).

# Symmetry-Driven Spin-Wave Gap Modulation in Nanolayered SrRuO₃/SrTiO₃ Heterostructures: Implications for Spintronic Applications

*Seung Gyo Jeong*[†,#], *Hyeonbeom Kim*[‡,#], *Sung Ju Hong*[‡,#], *Dongseok Suh*[‡,\*], *Woo Seok Choi*[†,\*]

[†]*Department of Physics, Sungkyunkwan University, Suwon 16419, Korea*

[‡]*Department of Energy Science, Sungkyunkwan University, Suwon 16419, Korea*

[\*]Corresponding authors: choiws@skku.edu; energy.suh@skku.edu

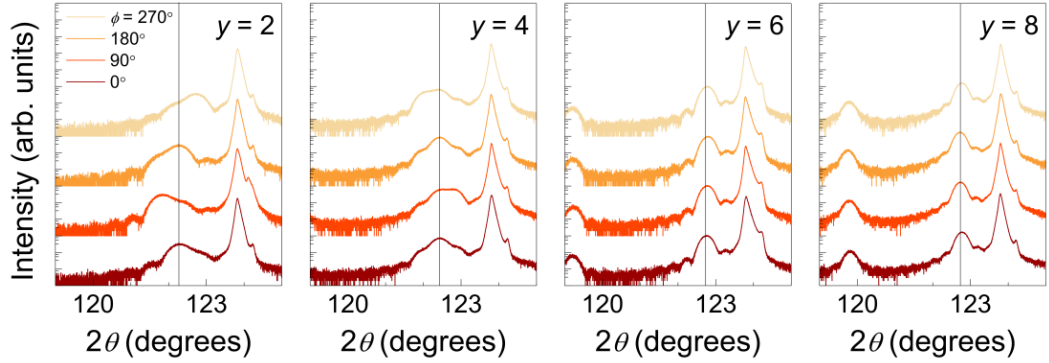

**Figure S1.** Off-axis XRD measurements of the [(SRO)$_6$|(STO)$_y$]$_{10}$ SLs with different *y* values (*y* = 2, 4, 6, and 8). The measurement was performed around the STO (204) Bragg reflections with $\phi$-angles of 0, 90, 180, and 270º. The vertical lines are guides to the eye.

**Supplemental Note 1.** $\rho_{xx}$ (*T*) curves of the SRO SLs with different *y* values.

Figure S2(a) shows the *T*-dependent $\rho_{xx}$ of the SRO SLs with a typical kink near the $T_c$ of ~130 K. The $\rho_{xx}$ (*T*) curves can be described by $\rho_{xx}$ (*T*) = $\rho_0$ + $AT^\alpha$, where $\rho_0$, *A*, and *α* are the residual resistivity, a coefficient, and the scaling parameter, respectively. Figure S3(b) shows the temperature derivative of the resistivity ($d\rho_{xx}$ (*T*) / d*T*) for the SRO SLs. In particular, *α* is a representative parameter for describing the electrical transport characteristics of SRO for three different temperature regimes (see Inset of Figure S2(b)).[32, 33] Above $T_c$, the extracted *α* (0.5) indicates the bad metallic behavior of SRO. Below $T_c$, with decreasing *T*, the *T*-dependent *α* exhibits a crossover from the non-Fermi liquid (*α* = 1.5) to the Fermi liquid (*α* = 2) phase around 30 K. These results indicate that the typical *T*-dependent metallic behavior of SRO is maintained below the atomically thin SRO thickness.

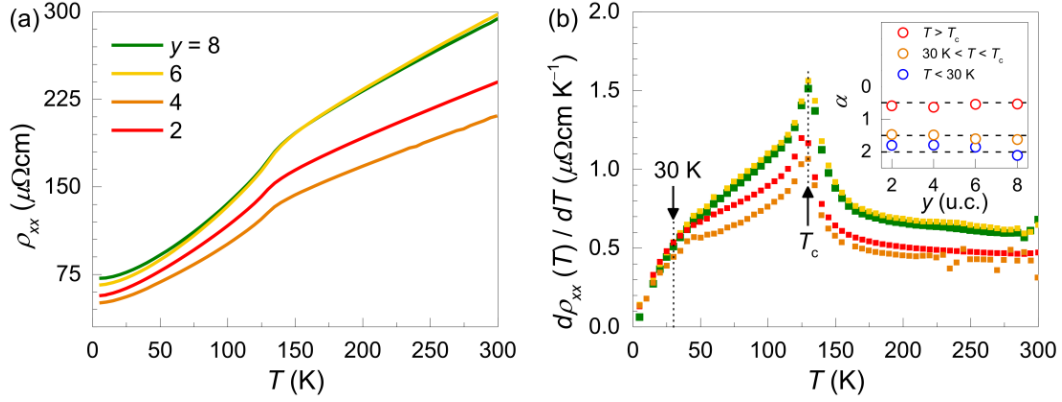

**Figure S2.** $T$-dependent $\rho_{xx}$ of the $[(SRO)_6|(STO)_y]_{10}$ SLs with different $y$ values. (a) The $\rho_{xx}(T)$ curves for all SLs consistently indicate metallic behavior, with a typical kink at $T_c$ (~130 K), which originates from the FM phase transition. (b) $d\rho_{xx}(T)/dT$ exhibits three regimes as a function of $T$. The inset in (b) shows the summarized $\alpha$ values of the SLs for different $y$ values. The dashed lines show the $\alpha$ values for the three phases of conventional SRO,[32, 33] *i.e.*, (1) $\alpha = 0.5$ ($T > T_c$, the bad metal), (2) $\alpha = 1.5$ (30 K $< T < T_c$, the non-Fermi liquid), and (3) $\alpha = 2$ ($T < 30$ K, the Fermi liquid).

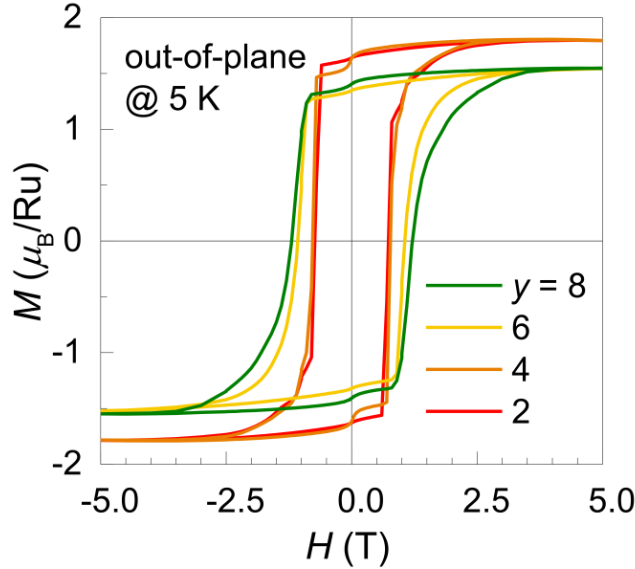

**Figure S3.** $M(H)$ curves of SRO/STO SLs. Out-of-plane $M(H)$ curves corresponding to $y = 2, 4, 6,$ and 8 SLs were measured at 5 K.

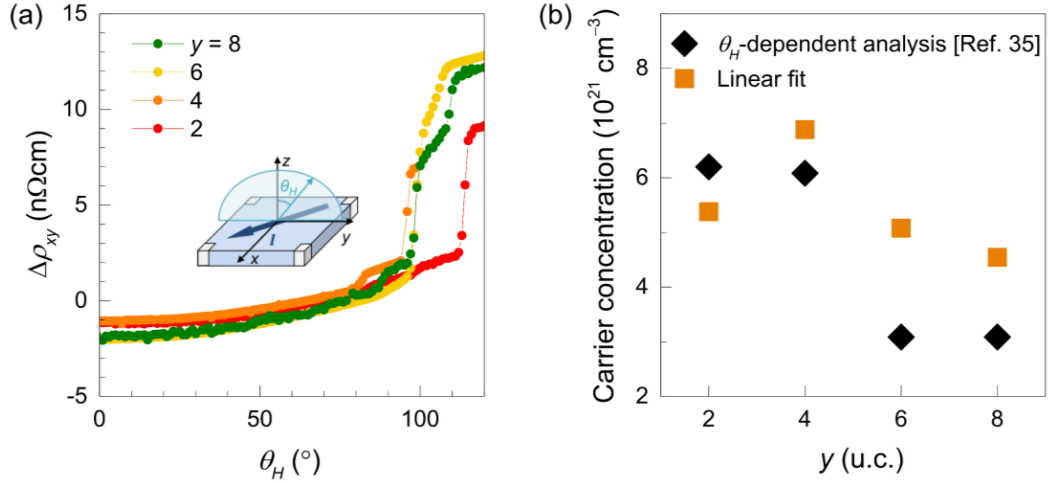

**Figure S4.** Separation of OHE and AHE in SRO/STO SLs using $\theta_H$-dependent analysis. (a) The $\Delta\rho_{xy}$ ($\rho_{xx}(\theta_H) - \rho_{xx}(0\,T)$) curves of SLs are shown as a function of $\theta_H$ at $H = 4$ T. The inset shows the schematic diagram of $\rho_{xy}(\theta_H)$ measurement configuration. (b) Comparison of carrier concentrations between the values obtained from the $\theta_H$-dependent analysis[35] and those from the conventional linear-fitting-based separation.

**Supplemental Note 2.** Structure-dependent carrier concentration and resultant $M_s$.

The carrier concentration of $SRO_o$ is larger than that of $SRO_t$. In principle, the carrier concentration of the nearly half-metallic SRO is sensitively determined by the down-spin state. An additional crystal field with regard to the tetragonal symmetry can reduce the down-spin state near the Fermi level.[S1] The enhanced carrier concentration of $SRO_o$ further increases $M_s$, owing to the increased number of spin-ordered itinerant carriers in $SRO_o$ compared to that in $SRO_t$. This result is consistent with those of previous SRO single-film studies.[S2]

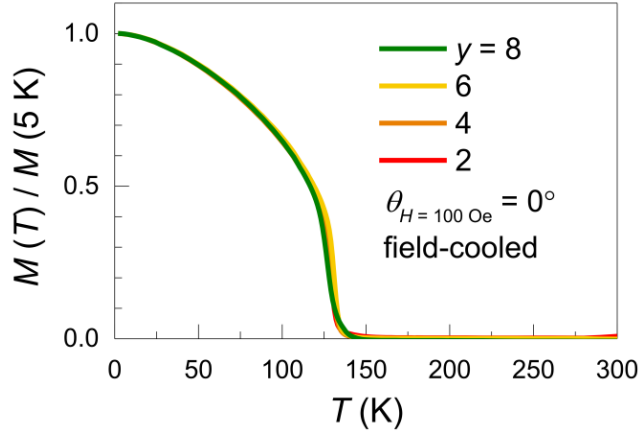

**Figure S5.** $M$ ($T$) curves of SRO/STO SLs. Field-cooled $M$ ($T$) curves of the SLs were measured with a 100 Oe magnetic field along the out-of-plane ($\theta_H = 0°$) direction. Each curve was normalized by the $M$ (5 K) values.

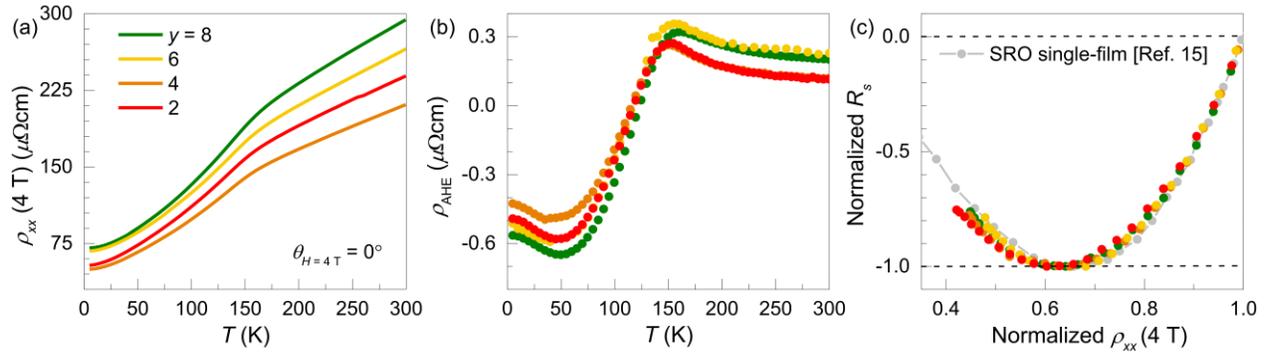

**Figure S6.** $T$-dependent magneto-transports and scaling behavior with $R_s$ ($\rho_{xx}$) for SRO/STO SLs. (a) The $\rho_{xx}$ ($T$) curves at $\theta_H = 0°$ at $H = 4$ T are shown. (b) We deliberately extracted $\rho_{AHE}$ (4 T) via utilizing the $\theta_H$-dependent analysis. (c) The normalized $R_s$ ($\rho_{xx}$) curves exhibit scaling behavior, consistent with those of SRO single films.[15]